\documentstyle[preprint,prd,aps,eqsecnum,epsf]{revtex}

\tightenlines

\begin{document}

\preprint{DFT-IF-UERJ-99/21}

\title{A Nonperturbative Study of Inverse Symmetry Breaking at High 
Temperatures}

\author{Marcus B. Pinto$^1\;$\thanks{Email address: fsc1mep@fsc.ufsc.br }
and Rudnei O. Ramos$^2\;$\thanks{Email address: rudnei@dft.if.uerj.br}} 

\address{
{\it $^1\;$ Departamento de F\'{\i}sica,}
{\it  Universidade Federal de Santa Catarina,}\\
{\it 88040-900 Florian\'{o}polis, SC, Brazil}\\
{\it $^2\;$ Departamento de F\'{\i}sica Te\'orica - Instituto de F\'{\i}sica,}
\\
{\it Universidade do Estado do Rio de Janeiro, }\\
{\it 20550-013 Rio de Janeiro, RJ, Brazil}}

\maketitle

\thispagestyle{empty}

\begin{abstract}

The optimized linear $\delta$-expansion is applied to multi-field 
$O(N_1) \times O(N_2)$ scalar
theories at high temperatures. Using the imaginary time formalism the
thermal masses are evaluated perturbatively up to order $\delta^2$ which
considers consistently all two-loop contributions. A variational 
procedure associated
with the method generates nonperturbative results which are used to
search for parameters values for inverse symmetry breaking (or symmetry
nonrestoration) at high temperatures. Our results are compared with the
ones obtained by the one-loop perturbative approximation, the gap
equation solutions and the renormalization group approach, showing good
agreement with the latter method. Apart from strongly supporting inverse
symmetry breaking (or symmetry nonrestoration), our results reveal the
possibility of other high temperature symmetry breaking patterns for which 
the last term in the breaking sequence is $O(N_1-1) \times O(N_2-1)$.

\vspace{0.34cm}
\noindent
PACS number(s): 11.10.Wx, 11.15.Tk, 98.80.Cq

\end{abstract}

\newpage

\setcounter{page}{1}

\section{Introduction}

The possibility that symmetries may be broken (or remain broken) at high
temperatures is not new \cite{weinberg}. This phenomenon is usually
called inverse symmetry breaking (ISB) (or symmetry nonrestoration
(SNR)). The idea of ISB (or SNR) is {\it per se} a very interesting one
due to its possible implementation in realistic particle physics models
and its consequences in the context of high temperature phase
transitions in the early Universe, with applications ranging from
problems involving CP violation and baryogenesis, topological defect
formation, inflation, etc (for a short list of the different
applications where SNR and ISB have been used see, {\it e.g.}, Refs.
\cite{appl1,appl2,appl3,appl4,appl5}). 

ISB or SNR are direct consequences that in two or multi-field theories
some of the coupling constants between fields can be negative while the
model is still bounded from below. This can be the case in any extension
of the standard model with a large scalar sector. Under these
conditions, there can be an enhanced symmetry breaking effect at high
temperatures, when thermal effects are taken into account in the model.
This effect is clear when one considers a one-loop analysis of the
effective potential in a simple model of two interacting scalar fields
or, as in  \cite{weinberg}, a $O(N) \times O(N)$ model.
However, this simple analysis is too naive for two main reasons: first,
at high temperatures the perturbative expansion in quantum field theory
becomes, in most cases unreliable. This happens because there can be
parameter regimes where powers of the coupling constants become
surmounted by powers of the temperature, or due to the appearance of
infrared divergences close to critical temperatures (as in field
theories displaying a second order phase transition or a weakly first
order transition). Second, the constrain conditions on the coupling
constants, which may be important for the observation of ISB/SNR,
usually requires large values for the couplings, in which case higher order
loop corrections may become important as well and change appreciably the
parameter space for ISB or SNR.

In order to account for the above problems of the one-loop
approximation, different methods have been used to analyze the question
of ISB in quantum field theory at high temperatures. The results,
however, have shown to be highly controversial, either by finding no
evidence for ISB/SNR or favoring the phenomenon. Examples of the former
were obtained in the context of methods like the large-N expansion
\cite{largeN1}, Gaussian effective potential \cite{gaussian}, chiral
Lagrangian technique \cite{chiral} and Monte Carlo simulations on the
lattice \cite{MC1}. Some applications which find evidence for ISB/SNR are 
Refs. \cite{gap,amel}, whose authors worked with self-consistent gap
equations; Ref. \cite{largeN2}, also in the context of the large-N expansion, 
but reaching a
different conclusion; 
Refs. \cite{r_group,pietro} in the context of the
renormalization group equations and Ref. \cite{MC2}, also in the context of
Monte Carlo simulations but this time supporting ISB/SNR. 

It is then obvious that it would be interesting to have a method
to clarify this question without involving the possible difficulties
related to the previous methods (like numerical precision in the Monte
Carlo simulations, self-consistency in the gap equations or resummed
perturbative methods, etc) used to study this problem. In this paper we
use a nonperturbative technique known as the linear $\delta$-expansion
(also known as optimized perturbation theory) \cite{linear,du} (for
earlier references see, {\it e.g.}, \cite{seznec}) and apply the method
to the $O(N_1) \times O(N_2)$ scalar model to obtain the thermal masses
to second order in the perturbative parameter $\delta$. We then
investigate their high temperature behavior to conclude about the
possibility of ISB/SNR. This model has already been considered before by
Bimonte and Lozano in \cite{gap} where SNR was studied by means of the
solutions of the gap equations of the model. By studying this theory we
can also extend our results to the simple model consisting of
two-interacting scalar fields, which has been extensively studied
before. In this case the symmetry group corresponds to $Z_2 \times Z_2$.
This step will allow us to compare our results with the ones furnished
by alternative methods. In a previous paper \cite{MR}, we have employed
the optimized linear $\delta$-expansion to study the resummation of
higher and leading order thermal corrections showing that the use of a
proper optimization scheme is equivalent to self-consistently solving
the gap equation for the thermal mass, where leading and higher order
infrared regularizing contributions are nonperturbatively taken into
account. An advantage of the linear $\delta$-expansion is that the same
simple propagator is used in the evaluation of any diagram, avoiding the
potential bookkeeping problems associated to other resummation methods.
Apart from being a powerful nonperturbative method, the optimized
perturbation theory was originally formulated as a general theory
applicable to arbitrary systems including strong interacting models,
which makes it particularly interesting to use in connection with ISB/SNR 
where the issue of large coupling constants is an important one.

This work is organized as follows. In Sec. II we introduce the model and
the lowest order one-loop result. In the same section, the linear
$\delta$-expansion technique is briefly described. It is then used, in
Sec. III, to evaluate the thermal masses up to order-$\delta^2$ in the
$3+1d$ model of interacting scalar fields with global $O(N_1) \times
O(N_2)$ symmetry. These calculations explicitly include two-loop
momentum independent as well as momentum dependent diagrams with equal
and different internal propagators. In Sec. IV we present our
optimization results for the thermal masses and investigate,
numerically, the possibility of ISB in the model. We specialize to the
case $N_1=90$ and $N_2=24$, where the model can be thought as
representing the Kibble-Higgs sector of a $SU(5)$ grand unified theory
and also to the case $N_1=N_2=1$, where it reduces to the $Z_2\times
Z_2$ model, which has been the object of many studies in connection with
ISB. Our predictions for the critical temperatures and size of the ISB
parameter region are compared with the ones found in the literature. In
Sec. V our concluding remarks are given. Two appendices are included to
present some technical details and for a brief discussion of
renormalization in the model.

\section{The Linear $\delta$-Expansion Applied to the
Evaluation of the Thermal Masses in the $O(N_1) \times O(N_2)$ Model}

\subsection{ The model and one-loop results} 

In this work we consider the scalar $O(N_1) \times
O(N_2)$ model described by 
\begin{equation} 
{\cal L} = \sum_{i=1}^{2}
\left [ \frac{1}{2} (\partial_{\mu}\phi_i)^2 - \frac{1}{2} m_i^2
\phi_i^2 - \frac {\lambda_i}{4!} \phi_i^4 \right ] -\frac {\lambda}{4}
\phi_1^2 \phi_2^2 +{\cal L}_{\rm ct}\;,
\label{model} 
\end{equation} 
where
\begin{equation} 
{\cal L}_{\rm ct}= \sum_{i=1}^{2} \left [ A_i
\frac{1}{2} (\partial_{\mu}\phi_i)^2 - \frac{1}{2} B_i \phi_i^2 - 
\frac{1}{4 !} C_i \phi_i^4
\right ] - \frac{1}{4} C \phi_1^2\phi_2^2 \;\;
\label{counter} 
\end{equation} 
represents the
counterterms needed to render the model finite. Note that ${\cal L}_{\rm
ct}$ requires an extra piece if one attempts to evaluate the thermal
effective potential \cite{Hatsuda}, which is not the case here. 
The boundness condition for the model described by (\ref{model})
requires that the coupling constants satisfy the inequalities

\begin{equation}
\lambda_1 > 0 , \;\; \lambda_2 > 0 \;\; {\rm and} \;\; 
\lambda_1 \lambda_2 > 9 \lambda^2 \;.
\label{bound}
\end{equation}

As noticed by Weinberg  \cite {weinberg} this boundness condition allows 
for negative values of the cross coupling $\lambda$ which may lead to 
the nonrestoration of symmetries at high temperatures. 
Another possibility is that theories which are symmetric at $T=0$ 
may be broken at large $T$ due to this negative value of $\lambda$. 
The thermal masses for this model have been first calculated using 
the one-loop approximation which, at high $T$, gives

\begin{equation}
M_1^2 \simeq m_1^2 + \frac{T^2}{24} \left [ \lambda_1 
\left ( \frac {N_1+2}{3} \right ) + \lambda N_2 \right ] \;,
\label{m1}
\end{equation}
and
\begin{equation}
M_2^2 \simeq m_2^2 + \frac{T^2}{24} \left [ \lambda_2 
\left ( \frac {N_2+2}{3} 
\right ) + \lambda N_1 \right ] \;.
\label{m2}
\end{equation}
Let us consider the interesting case where $\lambda < 0$ and  
setting $m_1^2$ and $m_2^2$ positive to have a symmetric theory at $T=0$. 
Inverse symmetry breaking takes place if one chooses, for example

\begin{equation}
\lambda > \frac{\lambda_1}{N_2}\left (\frac{N_1+2}{3} \right ) \,,
\end{equation}
which makes the $T^2$ coefficient of $M_1^2$ negative while the same 
coefficient for $M_2^2$ is kept positive, due to the boundness condition.
In this case, high temperatures will induce the breaking 
$O(N_1)\times O(N_2) \rightarrow O(N_1-1)\times O(N_2)$ 
at the critical temperature

\begin{equation}
\frac {T_c^2}{m_1^2} = 24 \left [ \lambda N_2 - 
\lambda_1 \left (\frac {N_1+2}{3} \right ) \right ]^{-1}\;.
\label{Tc1}
\end{equation}

\noindent
The one loop approximation results will be investigated 
numerically and compared to our results in Section IV.

\subsection {The interpolated model}

The optimized linear $\delta$-expansion is an
alternative nonperturbative approximation which has been
successfully used in  a plethora of different problems in  particle theory 
\cite {du,MR,Hatsuda,okotem,landau,njlft}, quantum mechanics \cite{ian,guida}, 
statistical physics \cite {alan}, nuclear matter 
\cite {gas} and lattice field theory \cite{evans}. One advantage of 
this method is that
the selection and evaluation (including renormalization) of {}Feynman
diagrams are done exactly as in ordinary perturbation theory using a
very simple modified propagator which depends on an arbitrary mass
parameter. Nonperturbative results are then obtained by fixing this
parameter.
The standard application of the linear
$\delta$-expansion to a theory described by a Lagrangian density 
${\cal L}$ starts with an interpolation defined by 
\begin{equation}
{\cal L}^{\delta} = (1-\delta){\cal L}_0(\eta) + \delta {\cal L} = 
{\cal L}_0(\eta) + \delta [{\cal L}-{\cal L}_0(\eta)],
\label{int}
\end{equation}

\noindent 
where ${\cal L}_0(\eta)$ is the Lagrangian density of a
solvable theory which can contain arbitrary mass parameters ($\eta$).
The Lagrangian density ${\cal L}^{\delta}$ interpolates between the
solvable ${\cal L}_0(\eta)$ (when $\delta=0$) and the original ${\cal
L}$ (when $\delta=1$). {}For the present model one may choose 

\begin{equation} 
{\cal L}_0(\eta_i) = \sum_{i=1}^{2} \left [ \frac{1}{2}
(\partial_{\mu}\phi_i)^2 - \frac{1}{2} m_i^2 \phi_i^2 - \frac{1}{2}
\eta_i^2 \phi_i^2 \right ] \;,
\end{equation}

\noindent 
and following the general prescription one can write

\begin{equation}
{\cal L}^{\delta} =\sum_{i=1}^{2}\left [ \frac{1}{2} 
(\partial_{\mu}\phi_i)^2 - \frac {1}{2} \Omega_i^2 \phi_i^2 
-\delta \frac {\lambda_i}{4!} \phi_i^4 + \frac {\delta}{2} 
\eta_i^2 \phi_i^2 \right ] - \delta \frac {\lambda}{4} \phi_1^2 
\phi_2^2 + {\cal L}^{\delta}_{\rm ct}\;,
\end{equation}

\noindent
where $\Omega_i^2=m_i^2+\eta_i^2$. The term ${\cal L}_{\rm ct}^{\delta}$, which contains the counterterms needed to render the model finite, is given by
\begin{equation}
{\cal L}_{\rm ct}^{\delta}=\sum_{i=1}^{2} \left [ A_i^{\delta}\frac{1}{2} 
(\partial_{\mu}\phi_i)^2 
-\frac{1}{2} B_i^{\delta} (\Omega_1,\Omega_2) \phi_i^2 - 
\frac{1}{4 !}\delta C_i^{\delta}\phi_i^4 + 
\frac{1}{2} \delta  B_i^{\delta}(\eta_1,\eta_2) 
\phi_i^2 \right ] - 
\frac{1}{4}\delta C^{\delta} \phi_1^2\phi_2^2\;,
\label{ctdelta}
\end{equation}

\noindent
where $A_i^\delta, \; B_i^{\delta},\; C_i^{\delta}$ and $C^{\delta}$
are the  counterterms coefficients.
One should note that the $\delta$-expansion interpolation introduces
only ``new" quadratic terms not altering the renormalizability of the
original theory. That is, the counterterms contained in ${\cal
L}^{\delta}_{\rm ct}$, as well as in the original ${\cal L}_{\rm ct}$,
have the same polynomial structure.

The general way the method works becomes clear by looking at the Feynman
rules generated by ${\cal L}^{\delta}$. First, the original $\phi_i^4$
vertex has its original Feynman rule $-i \lambda_i$ modified to $-i\delta
\lambda_i$ (the same applies to the mixed vertex $\phi_1^2\phi_2^2$). This minor modification is just a reminder that one is really
expanding in orders of the artificial parameter $\delta$. Most
importantly, let us look at the modifications implied by the addition of
the arbitrary quadratic part. The original bare propagator, 

\begin{equation}
S(k)= i(k^2-m_i^2 +i\epsilon)^{-1}\;, 
\end{equation}

\noindent
becomes
\begin{equation}
S(k)= i(k^2-\Omega_i^2 +i\epsilon)^{-1}=
{i \over {k^2 - m_i^2 + i\epsilon  }}\left[ 1 - 
{\frac{i}{k^2-m_i^2 +i\epsilon} (-i\eta_i^2)}
\right ]^{-1}\,,
\label{prop}
\end{equation}

\noindent
indicating that the term proportional to $ \eta_i^2 \phi_i^2$ contained in
${\cal L}_0$ is entering the theory in a nonperturbative way. On the
other hand, the piece proportional to $\delta\eta_i^2 \phi_i^2$ is only
being treated perturbatively as a quadratic vertex (of weight $i \delta
\eta_i^2$). Since only an infinite order calculation would be able to
compensate for the infinite number of ($-i\eta_i^2$) insertions contained
in Eq.~(\ref {prop}), one always ends up with a $\eta_i$ dependence in any
quantity calculated to finite order in $\delta$. Then, at the end of the
calculation one sets $\delta=1$ (the value at which the original theory
is retrieved) and fixes $\eta_i$ with the variational procedure known as
the Principle of Minimal Sensitivity (PMS) \cite {pms}

\begin{equation}
\frac{\partial P(\eta_i)}{\partial \{\eta_i\}} |_{\bar \{\eta_i\}} =0\;,
\end{equation}
where $P$ represents a physical quantity calculated 
{\it perturbatively} in 
powers of $\delta$ and then extremized with relation to
the $\eta_i$ parameters\footnote{{}For a discussion of convergence as
well as renormalization in the method, see 
Ref. \cite{MR} and references therein.}. This optimization
procedure applied to the thermal masses will be discussed 
in Sec. IV.

\section{The Thermal Masses up to order-$\delta^2$}

We can now start our evaluation of the thermal masses, defined by

\begin{equation}
M_i^2= \Omega_i^2 + \Sigma^{\delta}_i(p)\;,
\end{equation}

\noindent
where $\Sigma^{\delta}_i(p)$ is the thermal self-energy. 
At lowest order (first
order in $\delta$) the relevant contributions, which are momentum
independent, are given by ($i,j=1,2$ and $i\neq j$)

\begin{equation}
\Sigma_{i,1}^{\delta^1}(p)= - \delta \eta_i^2 + \delta 
\frac {\lambda_i}{2} \left ( \frac {N_i+2}{3} \right ) \int_T
\frac{d^d k}{(2\pi)^d} \frac {i}{k^2-\Omega_i^2+i\epsilon}
+\delta \frac {\lambda}{2} N_j \int_T
\frac{d^d k}{(2\pi)^d} \frac {i}{k^2-\Omega_j^2+i\epsilon}\;.
\label{S1a}
\end{equation}

\noindent
The temperature dependence can be readily obtained by using the 
standard imaginary time formalism prescription

\begin{equation}
p_0 \rightarrow i \omega_n \;\;, \;\;\;\;\;\; \int_T 
\frac{d^d k}{(2\pi)^d} \rightarrow i T \sum_n \int\frac{d^{d-1} {\bf
k}}{(2\pi)^{d-1}}\;.
\label{prescription}
\end{equation}
Then, the self-energy becomes

\begin{equation}
\Sigma^{\delta^1}_i(p) = - \delta \eta_i^2 + \delta  T \frac {\lambda_i}{2} 
\left ( \frac {N_i+2}{3} \right )
\sum_n \int 
\frac{d^{d-1} {\bf
k}}{(2\pi)^{d-1}} \frac {1}{\omega_n^2+E_i^2} +  \delta  T 
\frac {\lambda}{2} N_j
\sum_n \int 
\frac{d^{d-1} {\bf
k}}{(2\pi)^{d-1}} \frac {1}{\omega_n^2+E_j^2}\;,
\label{S1b}
\end{equation}

\noindent
where $E^2={\bf k}^2+\Omega^2$. Summing over Matsubara's frequencies one 
gets

\begin{eqnarray}
\Sigma^{\delta^1}_i(p)&=& - \delta \eta_i^2 + \delta \frac{\lambda_i}{2}
\left ( \frac {N_i+2}{3} \right )  
\int \frac{d^{d-1} {\bf
k}}{(2\pi)^{d-1}} \left \{ \frac {1}{2 E_i} - \frac {1}{E_i[
1 -\exp(E_i/T)]}\right \} \nonumber \\
&+& \delta \frac{\lambda}{2}N_j 
\int \frac{d^{d-1} {\bf
k}}{(2\pi)^{d-1}} \left \{ \frac {1}{2 E_j} - \frac {1}{E_j[
1 -\exp(E_j/T)]}\right \}\;.
\label{S1c}
\end{eqnarray}

\noindent
Then, using dimensional regularization \cite{ra} ($d=4-2 \epsilon$)
one obtains the thermal mass

\begin{eqnarray}
M_{i}^2 &=& \Omega_i^2 - \delta \eta_i^2 + 
\delta\frac{\lambda_i}{32 \pi^2}\left ( \frac {N_i+2}{3} \right )
\left \{ \Omega_i^2  
\left [ - \frac {1}{\epsilon} + 
\ln \left (\frac{\Omega_i^2}{4 \pi \mu^2} \right)+
\gamma_E-1
\right] + 32 \pi^2 T^2 
h\left(\frac{\Omega_i}{T}\right) \right \} \nonumber \\
&+& \delta\frac{\lambda}{32 \pi^2}N_j \left \{ \Omega_j^2  
\left [ - \frac {1}{\epsilon} + 
\ln \left (\frac{\Omega_j^2}{4 \pi \mu^2} \right)+
\gamma_E-1
\right] + 32 \pi^2 T^2  
h\left(\frac{\Omega_j}{T}\right) \right \}\;,
\label{S1}
\end{eqnarray}
where $\mu$ is a mass scale introduced by dimensional regularization and  
\begin{equation}
h(y_i) = \frac {1}{4 \pi^2}  \int_0^{\infty} dx \frac
{x^2}{[x^2+y_i^2]^{\frac{1}{2}}[ \exp(x^2+y_i^2)^{\frac{1}{2}}-1] }\;.
\label{hyint}
\end{equation}

\noindent
Note that the temperature independent term diverges and must be 
renormalized. In this paper we chose the Minimal Subtraction (MS) 
scheme where the counterterms eliminate the poles only. At this order 
the only divergence in (\ref{S1}) is
\begin{equation}
\Sigma^{\delta^1}_{\rm div,i} = -\frac{\delta}{32 \pi^2 \epsilon}
\left(\lambda_i \frac {N_i+2}{3}\Omega_i^2 + 
\lambda N_j \Omega_j^2 \right)\;,
\label{Sdiv1}
\end{equation}
which is easily eliminated by the O($\delta$) mass 
counterterm 
\begin{equation}
\Sigma^{\delta^1}_{\rm ct,i} =  B^{\delta^1}_i (\Omega_1,\Omega_2) 
=\frac{\delta}{32 \pi^2 \epsilon}
\left(\lambda_i \frac {N_i+2}{3}\Omega_i^2 + 
\lambda N_j \Omega_j^2 \right) \;.
\label{ct1}
\end{equation}

\noindent
By looking at Eq. (\ref{S1}) one can see that the terms proportional 
to $\delta$ represent exactly those that appear 
at first order in the coupling constants in ordinary perturbation theory excepted that 
we now have $\Omega^2_i$ instead of $m^2_i$, 
$\delta \lambda_i$ instead 
of $\lambda_i$ and $\delta \lambda$ instead of $\lambda$. Therefore, 
it is not surprising that to this order the 
renormalization procedure implied by the interpolated theory is 
identical to the procedure implied by the original theory at
first order in the coupling constants. 

Let us now analyze the temperature dependent integral which is expressed, 
in the
high temperature limit ($y_i=\Omega_i/T \ll 1$), as \cite{Kapusta}
\begin{equation}
h(y_i) =
\frac {1}{24} - \frac{1}{8\pi} y_i
- \frac{1}{16\pi^2} y_i^2 \left[ \ln \left ( \frac
{y_i}{
4\pi} \right ) + \gamma_E - \frac{1}{2} \right ] + {\cal O}(y_i^3)\;\;.
\label{hy}
\end{equation}
In principle, since $\eta_i$ is arbitrary, one could be reluctant in 
taking the limit $\Omega_i/T \ll 1$. However, as discussed 
in our previous work \cite{MR}, the 
use of both forms for 
the integral $h(y_i)$  does not lead
to any significant numerical changes in the optimization procedure. 
This is also true in the present work.
Then, by taking $h(y_i)$ in the high temperature limit, 
one obtains the ${\cal O}(\delta)$ thermal mass: 

\begin{equation}
M_i^2 = \Omega_i^2 - \delta \eta_i^2 +\delta \lambda_i 
\left ( \frac {N_i+2}{3}\right ) X_i(T)+\delta \lambda N_j X_j(T)
+{\cal O}(\delta^2)\;,
\label{Mi1}
\end{equation}
where we have defined the quantity $X_i (T)$ 

\begin{equation}
X_i(T)= \frac{T^2}{24}-
\frac{ T\Omega_i}{8\pi} + 
\frac{ \Omega_i^2 }{32\pi^2} L(T) 
\;\;,
\end{equation}
with $L(T)$ given by
\begin{equation}
L(T)=\ln\left(\frac{4 \pi T^2}{\mu^2}\right ) - \gamma_E \;.
\end{equation}

\noindent
{}For notational convenience when expressing the remaining
contributions to the self-energies, we also define the additional 
quantities $Y_i(T),
\; Z_i(0), \; W_i (0)$ and $R_i(T)$ given by

\begin{equation}
Y_i(T)= - \frac{ T\Omega_i}{16\pi} + 
\frac{ \Omega_i^2 }{32\pi^2} L(T) \;,
\end{equation}

\begin{equation}
Z_i(0) =\frac {1}{2}\left[ \ln \left (
\frac{\Omega_i^2}{4\pi \mu^2} \right) +\gamma_E \right]^2 + 
\frac{\pi^2}{12} \;,
\end{equation}

\begin{equation}
W_i(0)=\frac {1}{2} \left[ \ln \left (
\frac{\Omega_i^2}{4\pi \mu^2} \right ) + \gamma_E -1 \right]^2 +\frac{1}{2}
+ \frac{\pi^2}{12}\;, 
\end{equation}

\noindent
and

\begin{equation}
R_i(T)=- \frac{ T\Omega_i^2}{16\pi \Omega_j} + 
\frac{ \Omega_i^2 }{32\pi^2} L(T) \;.
\end{equation}

At ${\cal O}(\delta^2)$ the self-energy receives contributions from
momentum independent as well as momentum dependent diagrams. Let us
first consider the momentum independent diagrams, which to this order
have one and two loops. The order-$\delta^2$, one loop, momentum
independent contribution is given (in the high temperature limit) by 

\begin{equation}
\Sigma^{\delta^2}_{i,2} = \delta^2 \lambda_i \eta_i^2
\left ( \frac {N_i+2}{3}\right ) \left [\frac{1}{32\pi^2 \epsilon}- 
\frac{Y_i (T)}{\Omega_i^2} \right ] \;,
\label{S2}
\end{equation}

\noindent
and
\begin{equation}
\Sigma^{\delta^2}_{i,3} = \delta^2 \lambda \eta_j^2
N_j\left [ \frac{1}{32\pi^2 \epsilon}- \frac{Y_j (T)}{\Omega_j^2} 
\right ] \;\;\;.
\label{S3}
\end{equation}

\noindent
As discussed in Appendix B, the above contributions can be rendered finite 
using  the
mass type counterterms 
contained in ${\cal L}^{\delta}_{\rm ct}$ which are tailored to account for 
divergences arising from the extra quadratic vertices introduced during 
the interpolation process. The momentum independent two loop contribution 
is given by the four ``double scoop" diagrams

\begin{eqnarray}
\Sigma^{\delta^2}_{i,4} &=&
\delta^2 \lambda_i^2 \left ( \frac {N_i+2}{3} \right )^2\left \{
\frac {\Omega_i^2}{(32\pi^2)^2 \epsilon^2}  -
\frac{1}{32\pi^2 \epsilon} 
\left [ X_i(T) + Y_i(T)
\right ] 
-\frac{T^3}{384 \pi \Omega_i}+\frac{T^2}{128\pi^2} \right.
\nonumber \\
&+&\left. \frac{  L(T)}{(16\pi)^2}\left [ 8 X_i(T) - 
\frac{T \Omega_i}{2 \pi} \right ]
+\frac {  \Omega_i^2}{(32 \pi^2)^2}\left [ 
Z_i(0) + W_i(0) \right ]  \right \}\;,
\label{S4}
\end{eqnarray}

\begin{eqnarray}
\Sigma^{\delta^2}_{i,5} &=&
\delta^2 \lambda \lambda_i \frac {(N_i+2)}{3} N_j \left \{ 
\frac {\Omega_j^2}{(32\pi^2)^2 \epsilon^2}   -
 \frac{1 }{32\pi^2 \epsilon}  
\left [ X_j(T) + R_j(T) \right ] 
-  \frac{T^3}{384 \pi \Omega_i}+\frac{T^2}{128\pi^2} 
\frac{\Omega_j}{\Omega_i}\right.
\nonumber \\
&+&\left. \frac{L(T)}{(16\pi)^2}\left [ 8 X_j(T) - 
\frac {T \Omega_j^2}{2 \pi \Omega_i}
\right ]
+\frac{ \Omega_j^2}{(32 \pi^2)^2}\left[ Z_i(0) + W_j(0)
\right ] \right \} \;,
\label{S5}
\end{eqnarray}

\begin{eqnarray}
\Sigma^{\delta^2}_{i,6}&=&\delta^2 \lambda \lambda_j 
\frac {(N_j+2)}{3} N_i\left \{
\frac {\Omega_j^2}{(32\pi^2)^2 \epsilon^2}  -
\frac{1}{32\pi^2 \epsilon} 
\left [ X_j(T) + Y_j(T)
\right ] 
-\frac{T^3}{384 \pi \Omega_j}+\frac{T^2}{128\pi^2} \right.
\nonumber \\
&+&\left. 
\frac{ L(T)}{(16\pi)^2}\left [ 8 X_j(T) - \frac{T \Omega_j}{2 \pi} \right ]
+\frac {  \Omega_j^2}{(32 \pi^2)^2}\left [ 
Z_j(0) + W_j(0) \right ] \right \}
\label{S6} 
\end{eqnarray}

\noindent
and

\begin{eqnarray}
\Sigma^{\delta^2}_{i,7}&=&\delta^2 \lambda^2  N_i N_j \left \{ 
\frac {\Omega_i^2}{(32\pi^2)^2 \epsilon^2}   -
 \frac{1 }{32\pi^2 \epsilon}  
\left [ X_i(T) + R_i(T) \right ] 
-  \frac{T^3}{384 \pi \Omega_j}+\frac{T^2}{128\pi^2} 
\frac{\Omega_i}{\Omega_j} \right.
\nonumber \\
&+&\left. \frac{L(T)}{(16\pi)^2}
\left [ 8 X_i(T) - \frac {T \Omega_i^2}{2 \pi \Omega_j}
\right ]
+\frac { \Omega_i^2}{(32 \pi^2)^2}\left[ Z_j(0) + W_i(0)
\right] \right \}\;.
\label{S7} 
\end{eqnarray}

\noindent
To render these diagrams finite one needs  mass and vertex counterterms 
\cite{ra}. Considering the ${\cal O}(\delta)$ mass counterterms used 
to eliminate 
the divergences in $\Sigma_{\rm div,i}^{\delta^1}$ (see Eq. (\ref{ct1})), 
one is able to
build two one loop ${\cal O}(\delta^2)$ diagrams whose contributions 
are given by

\begin{equation}
\Sigma^{\delta^2}_{i,8} = \frac{\delta^2 \lambda_i (N_i+2)}{3(32 \pi^2)^2} 
\left[ \frac{\lambda_i \Omega_i^2
(N_i+2)}{3} + \lambda \Omega_j^2 N_j \right] 
\left\{ - \frac{1}{\epsilon^2} +
\frac{32 \pi^2}{\Omega_i^2} Y_i (T)  \frac{1}{\epsilon} - Z_i (0) \right\}\;,
\label{S8} 
\end{equation}

\noindent
and

\begin{equation}
\Sigma^{\delta^2}_{i,9} = \frac{\delta^2 \lambda N_j}{(32 \pi^2)^2} 
\left[ \frac{\lambda_j \Omega_j^2
(N_j+2)}{3} + \lambda \Omega_i^2 N_i \right] 
\left\{ - \frac{1}{\epsilon^2} + 
\frac{32 \pi^2}{\Omega_j^2} Y_j (T) \frac{1}{\epsilon} - Z_j (0) \right\}\;.
\label{S9} 
\end{equation}

\noindent
Additionally, from the vertex counterterms appearing in (\ref{ctdelta}),
with $C_i$ and $C$ (at order ${\cal O}(\delta^2)$) given by

\begin{equation}
C_i = \frac{\delta^2}{32 \pi^2 \epsilon}\left[
\lambda_i^2 \frac{(N_i+8)}{3} + 3 \lambda^2 
N_j \right] \;,
\label{Ci}
\end{equation}
and 
\noindent
\begin{equation}
C = \frac{2 \delta^2 \lambda^2}{16 \pi^2 \epsilon} + 
\frac{\delta^2 \lambda}{96 \pi^2 \epsilon} \left[ \lambda_1
(N_1 +2) + \lambda_2 (N_2+2) \right]\;,
\label{C}
\end{equation}

\noindent
one can built the one-loop ${\cal O}(\delta^2)$ diagrams:

\begin{equation} 
\Sigma^{\delta^2}_{i,10} = - \frac{\delta^2}{16 \pi^2}\frac{(N_i+2)}{3} 
\left [ \lambda_i^2  
\frac{(N_i+8)}{6} +
\lambda^2 \frac {3 N_j}{2} \right ]
\left [  \frac{\Omega^2_i}{32 \pi^2 \epsilon^2} -
\frac{1}{\epsilon} X_i(T) + \frac {\Omega^2_i}{32 \pi^2} W_i(0) 
\right] \;, 
\label{S10}
\end{equation}

\noindent
and

\begin{eqnarray} 
\Sigma^{\delta^2}_{i,11} &=&- \frac {\delta^2 N_j}{32 \pi^2} 
\left \{ 4\lambda^2+ 
\frac{\lambda}{3} \left[ \lambda_i
(N_i +2) + \lambda_j (N_j+2) \right]  \right \} \nonumber\\
&\times&\left [  \frac{\Omega_j^2}{32 \pi^2 \epsilon^2} - 
\frac{1}{\epsilon} X_j(T) +\frac{\Omega_j^2}{32 \pi^2 } W_j(0) 
\right ]\;.
\label{S11}
\end{eqnarray}

The next contribution to the self-energy at ${\cal O}(\delta^2)$ 
comes from the two-loop ``setting sun"
diagrams. These momentum dependent contributions are given by 
the setting sun diagram with equal mass internal propagators, 

\begin{equation}
\Sigma_{i,12}^{\delta^2}(p)= -\delta^2 \frac{\lambda_i^2 (N_i+2)}{18}  
\left(G_{iii,0}
+ G_{iii,1} + G_{iii,2} \right) \;,
\label{S12a}
\end{equation}
and by the one with internal propagators with different masses,
\begin{equation}
\Sigma_{i,13}^{\delta^2}(p)= -\delta^2 \frac{\lambda^2  N_j}{2}  
\left(G_{ijj,0}
+ G_{ijj,1} + G_{ijj,2} \right) \;.
\label{S13a}
\end{equation}

\noindent
In the above expressions,
$G_0$ is the zero temperature part 
of the diagrams and $G_1$ and
$G_2$ are the finite temperature ones (with one and two Bose factors,
respectively).
The details of the evaluation of (\ref{S12a}) and (\ref{S13a})
are given in Appendix A. Their contributions are

\begin{eqnarray}
{\rm Re}[\Sigma^{\delta^2}_{i,12}(p)] &=& 
\delta^2 \frac {\lambda_i^2 (N_i+2)}{3 (32 \pi^2)^2} \left [
\frac {\Omega_i^2}{\epsilon^2} +
\frac {\Omega_i^2}{\epsilon} - 
\frac {p^2}{6 \epsilon} 
- \frac{64 \pi^2 X_i(T)}{\epsilon} \right ] \nonumber \\
&+&\delta^2 \frac{\lambda_i^2 (N_i+2) \Omega_i^2}{6 (4 \pi)^4} 
\left[ \ln^2 \left( \frac{
\Omega_i^2}{4 \pi \mu^2} \right) + \left(2 \gamma_E - \frac{17}{6} \right) 
\ln\left(
\frac{\Omega_i^2}{4 \pi \mu^2} \right) + 1.9785 \right ]\nonumber \\
&+& \delta^2 \frac {\lambda_i^2 (N_i+2)}{3 (4 \pi)^2} \left[ \ln \left ( 
\frac {\Omega_i^2}{4 \pi \mu^2} \right ) - 2 + \gamma_E \right ] \nonumber \\
&\times& \left \{  \frac {T^2}{24} - \frac {T \Omega_i}{ 8 \pi} 
- \frac {\Omega_i^2}{16 \pi^2} \left [ \ln \left (
\frac {\Omega_i}{4 \pi T}\right ) +\gamma_E  - \frac{1}{2}  \right ] 
\right \} \nonumber \\
&+& \delta^2 \frac {\lambda_i^2 (N_i+2) T^2}{72 (4 \pi)^2}\left [ \ln \left ( 
\frac {\Omega_i^2}{T^2} \right ) + 5.0669 \right ] \;,
\label{S12} 
\end{eqnarray}

\noindent
and 

\begin{eqnarray}
\lefteqn{{\rm Re}[\Sigma^{\delta^2}_{i,13}(p)]  = 
\delta^2 \frac{\lambda^2  N_j}{2}
\frac{\Omega_j^2}{(4\pi)^4} 
\left \{ \frac{1}{\epsilon^2}
\left (1 +\frac {n^2}{2} \right ) 
+ \frac {1}{\epsilon} 
\left ( 1+\frac {n^2}{2} \right ) \left [ 3 - 2 \gamma_E - 2 \ln 
\left ( \frac {\Omega_j^2}{4\pi \mu^2} \right ) \right ] \right.}
\nonumber \\
& & \left.
- \frac{p^2}{4 \Omega_j^2 \epsilon} - \frac{n^2}{\epsilon} \ln (n^2)  
\right \} 
- \delta^2 N_j \frac{\lambda^2 T^2}{ (4\pi)^2} 
\frac{1}{\epsilon}\left[ h\left(\frac{\Omega_i}{T}\right) +
2 h\left(\frac{\Omega_j}{T}\right) \right] \nonumber \\
& &+ \delta^2 \frac{\lambda^2  N_j}{2}
\frac{\Omega_j^2}{(4\pi)^4}  \left( 1 +\frac {n^2}{2} \right ) 
\left [ 7 + \frac{\pi^2}{6} - 6 \gamma_E + 2 \gamma^2_E - 2(3 - 2 \gamma_E)
\ln \left ( \frac {\Omega_j^2}{4\pi \mu^2} \right ) + 2 \ln^2 \left ( 
\frac {\Omega_j^2}{4\pi \mu^2} \right ) \right ] \nonumber \\
& &- \delta^2 \frac{\lambda^2  N_j}{2}
\frac{\Omega_j^2}{(4\pi)^4}\left [ 1 + \frac {11}{8}n^2 - \left ( 1 + 
\frac{n^2}{2} \right ) \ln(n^2) - \frac{1}{2} n^2 \ln^2(n^2) 
+ \frac {(1-n^2)^2}{n^2} \left ( {\rm Li}_2(1-n^2) - \frac{\pi^2}{6} 
\right ) \right ]
\nonumber \\
& &- \delta^2 \frac{\lambda^2  N_j}{2}
\frac{\Omega_i^2}{(4\pi)^4}\left \{ \left [ 3 - 2 \gamma_E - 
2 \ln\left ( \frac {\Omega_j^2}{4\pi \mu^2} \right ) \right ] 
\left ( \frac {1}{4} + \ln (n^2) \right ) \right \}
\nonumber \\
& &+ 
 \delta^2 N_j \frac{\lambda^2 T^2}{ (4 \pi)^2} 
\left\{
h\left(  \frac{\Omega_i}{T}\right)
\left [ \ln\left(
\frac{
\Omega_j^2}{4 \pi \mu^2} \right) -2 + \gamma_E \right ]
+ 2 h\left(  \frac{\Omega_j}{T}\right)
\left [ \ln\left(
\frac{
\Omega_i^2}{4 \pi \mu^2} \right) -2 + \gamma_E \right ] \right\}
\nonumber \\
& & 
+ \delta^2 N_j \frac{\lambda^2 T^2}{8 (4 \pi)^2}
\left[ 2 \ln \left( \frac{\Omega_i + 2 \Omega_j}{3 T}\right) + 5.0669 \right]
\;.
\label{S13}
\end{eqnarray}
where $n=\Omega_i/\Omega_j$.

{}From Eqs. 
(\ref{S1}), (\ref{S2})-(\ref{S9}), (\ref{S10}), (\ref{S11}),
(\ref{S12}) and (\ref{S13})
one
easily sees that all the temperature dependent divergences cancel 
exactly. The remaining divergences are handled in the usual way,
being canceled by the counterterms appearing in ${\cal L}_{\rm ct}$,
Eq. (\ref{counter}) (see Appendix B). The sum of the remaining 
finite terms of each contribution to the self-energy makes
the total contribution to the thermal mass up to
order $\delta^2$, $M^2_i = \Omega_i^2 - \delta \eta^2_i +
\Sigma_{i,1} + \ldots + \Sigma_{i,13}$.

\section{Numerical Results}

We are now in position to set $\delta=1$ and apply the PMS to the thermal 
masses. Before doing that few points concerning the optimization procedure 
should be clarified. {}First, we recall that our interpolation 
procedure has been carried out in a very general way by assigning 
a different interpolation parameter to each field. 
Although general, this procedure brings in two arbitrary parameters 
which have to be fixed with the PMS. Of course this complicates the 
numerical optimization procedure, which is needed at higher orders, 
since one has to look for extrema in the $\eta_1$, $\eta_2$ space. 
Sometimes these extrema show up as saddle points which are hard to detect 
numerically. In principle, bearing in mind that $\phi_1$ and $\phi_2$ 
are both the same type of fields, one could be tempted to use the freedom allowed by 
the interpolating process to set $\eta_1=\eta_2=\eta$. In order to 
assess the validity of such choice we have done the optimization in 
both ways finding that the case $\eta_1 \ne \eta_2$ 
gives better results.  The second point regards the actual quantity to 
be extremized. By looking at our equations one can easily see that, 
except for the case where $N_1=N_2$, $m_1^2=m_2^2$ and $\lambda_1=\lambda_2$, 
the PMS applied separately to $M_1^2$ and $M_2^2$, at the same temperature, 
can generate different values for the same $\bar \eta_i$. One way to 
avoid this would be to evaluate and optimize a more comprehensive 
quantity such as the effective potential. In fact this procedure has 
been advocated in other applications of the optimized $\delta$-expansion 
\cite {gas}. However, in the present work we have not attempted to 
evaluate the effective potential since, at ${\cal O}(\delta^2)$, this 
would imply in the evaluation of three-loop zero point functions 
which become rather cumbersome at finite temperatures
\footnote {See Ref. \cite {sunil} for a discussion concerning the 
$\delta$-expansion evaluation of the effective potential.}. 
Although this may be regarded as a controversial point within the method 
it will not  be discussed any further in the present application. 
Here we follow 
the original prescription given in \cite {pms} where it is suggested 
that the PMS should be applied to each different physical quantity so 
that $\eta_i$ can be adjusted to the relevant energy scale. The validity of such procedure will be judged by comparing our results with well known predictions in the $N_1=N_2=1$ limit.
{}Finally, as discussed in \cite {MR}, 
one can easily see that applying the PMS to $M_i^2$ at first order 
in $\delta$ produces coupling independent values for $\bar \eta_i$ which 
do not furnish truly nonperturbative results. Therefore we will only 
investigate the results generated at order-$\delta^2$.

We work in units of the arbitrary scale $\mu$ introduced by dimensional 
regularization, by extremizing the dimensionless quantity $M_i^2/\mu^2$. 
To search for inverse symmetry breaking at high temperatures one must 
ensure  that the symmetries are restored at $T=0$. This is achieved by 
setting $m_i^2 > 0$ and by observing the boundness condition while 
setting $\lambda \rightarrow - \lambda$ in all our equations. 

\subsection{The $Z_2 \times Z_2$ case.}

Let us start by studying the $N_1=N_2=1$ case where the theory reduces
to the $Z_2\times Z_2$ model which has been extensively studied in the
literature. In \cite {r_group} (see also \cite{pietro}) it is claimed
that the critical temperature is not reliably estimated by the loop
expansion which does not include temperature effects in the coupling
constants. On the other hand, the nonperturbative renormalization group
approach (RGA) used in \cite {r_group} does include those effects and as
a first check of the validity of our results we estimate the critical
temperature for the $Z_2 \times Z_2$ model comparing our results at
${\cal O}(\delta^2)$ with the ones furnished by the two methods
mentioned above. This is done in the first two rows of Table I and also
in {}Fig. 1. In all cases $m_1^2/\mu^2=m_2^2/\mu^2 = 1.0$. Just above
the critical temperatures presented in the first two rows the thermal
mass $M_1^2$ is negative whereas $M_2^2$ is positive. Our numerical optimization strategy is the following: for a temperature below the value
predicted by the 1-loop approximation (1LA) we identify the extremum in
the $\eta_1,\; \eta_2$ space. Both masses in the extremum positions are
found to be consistently positive and the extremum positions are found
to be unique (and consistent with the high temperature approximation, in
the sense that ${\bar \eta}_{1,2} < T$). The extremum is then followed as
the temperature is increased. When one of the masses becomes negative the
value of $T_c$ is obtained and the values of ${\bar \eta}_1, \;
{\bar \eta}_2$ are registered. {}For example, around the critical
temperature the values obtained by extremizing $M_1^2$ are $\bar \eta_1
\approx 0$, $\bar \eta_2 \approx 0.25 \; T$ and $\bar \eta_1 \approx 0$,
$\bar \eta_2 \approx 0.15 \; T$ for the parameters in the first and
second rows, respectively, in Table I. {}From this table one can see the
remarkable agreement between our results and the ones provided by the
RGA. In {}Fig. 1 we compare the numerical values of the critical
temperature with the ones predicted by the 1LA and RGA for several
values of the cross coupling. Once again our results agree with the ones
given by the RGA \cite {r_group,pietro}. We have also investigated the
behavior of the critical temperature as a function of $\lambda_2$ for
fixed $m_{1,2}^2/\mu^2 = 1.0$, $\lambda_1 = 0.009$ and $\lambda=
-0.025$. As can be seen directly from Eq. (\ref{m1}), the one-loop
predicts that $T_c/m_1 \sim 49.0$ for any value of $\lambda_2$, while
the RGA predicts that it increases with the latter. Once again our
results, not shown, agree with the ones given by the RGA. These findings
give us confidence about the correctness of the optimization procedure
adopted. 

The critical temperatures predicted by the $\delta$-expansion, as well
as by the RGA, are higher than those predicted by the 1LA. This could be
roughly understood by recalling that for the phase transitions
considered the PMS applied to the mass which signals the phase
transition ($M_1^2$ for the parameters of Table I) generates a ${\bar
\eta}_2$ which is an increasing function of $T$ whereas it gives ${\bar
\eta}_1 \approx 0$. Therefore, at high $T$, the optimized mass ${\bar
\Omega}_2$ which ``dresses" the $\delta$-expansion propagator associated
with $\phi_2$ is greater than its counterpart ${\bar \Omega}_1^2$. In
$M_1^2$, the quantity ${\bar \Omega}_2^2$ is always associated with the
coupling $\lambda$ which drives ISB. It seems that the increasing of
${\bar \Omega}_2^2$ with the temperature suppresses the effect of
$\lambda$ in enhancing ISB. Another interesting point to be discussed
regards the inclusion of temperature dependent coupling constants. While
this possible dependence is overlooked in the 1LA it is taken into
account in the RGA. These effects are more subtle to be observed
directly in our case. However, by looking at the nature of the diagrams
considered by us at order-$\delta^2$, one may have an idea of how these
effects enter our calculations. Basically, one can think of double
scoops and setting suns diagrams as being tadpole diagrams with vertex
corrections. Expanding the vertex to order-$\delta$ gives the simple one
loop diagrams considered in Eq. (\ref{Mi1}), while expanding the vertex
to order-$\delta^2$ gives the double scoops and setting suns which
indirectly contain one-loop temperature dependent corrections to the
vertex.

\subsection {The $O(N_1) \times O(N_2)$ case.}

Let us now consider the more realistic $O(N_1)\times O(N_2)$ case. We follow
Bimonte and Lozano, in the first reference in \cite{gap}, by choosing
$N_1 =90$ and $N_2 =24$ so that the model can be thought of as
representing the Kibble-Higgs sector of a $SU(5)$ grand unified model.
Let us start by estimating some critical temperatures and comparing our
results with the lowest order one-loop results given by Eqs. (\ref{m1})
and (\ref{m2}). The two first rows of Table II display our results for
the set of parameters considered in the previous subsection. Our
predicted values for the critical temperatures are proportionally much
higher than for the $N_1=N_2=1$ case. The two last rows of Table II
display the results for more typical parameter values for this model,
which are usually taken around $\lambda_1 = 0.8$ and $\lambda_2 = 1.0$
(see \cite {appl3}). These results suggest that the parameter region for
ISB predicted by the $\delta$-expansion is smaller than the one
predicted by the lowest order 1LA prediction. To illustrate that we
offer {}Figs. 2 and 3. {}Figure 2 shows our ${\cal O}(\delta^2)$ result
(dot-dashed line) compared with the results produced by the 1LA at
lowest order (thin continuous line) for fixed $\lambda_1=0.8$ and $T=5.0
\; \mu$. The upper parabola represents the limiting region for
boundness. The dashed line is the Bimonte and Lozano's next to leading
order one-loop (BLA) result for an arbitrarily large $T$. {}Fig. 3 is
similar to {}Fig. 2 except that $\lambda_1$ varies while $\lambda_2$ is
kept fixed at the unity value. Note that the next to leading order
correction considered in \cite{gap} for the one-loop approximation
reduces significantly the region of ISB given at lowest order. However,
the next to leading order calculation of the BLA approach considers up
to the second term, which is mass dependent, in the expansion of the
temperature dependent integral given by Eq. (\ref{hyint}). This means that their
procedure does not take into account logarithmic terms in the expansion
of $h(y_i)$ since they would give corrections of order $\lambda_i^2 \ln
\lambda_i$ which also arise from two-loop diagrams. Our calculations, on
the other hand, avoid these bookkeeping problems and all contributions
are consistently and explicitly considered in a given order. Moreover,
support for ISB/SNR also arises in the work of Amelino-Camelia 
Ref. \cite{amel} who, like us,  
considers two-loop diagrams in the simple $Z_2 \times Z_2$ model.

To illustrate how the temperature affects the ISB region we offer {}Fig.
4 where we have fixed $\lambda_1=0.8$ while $\lambda_2$ varies. It is
clear that the ISB parameter region for $T= 5.0 \; \mu$ is smaller than
for $T=10.0 \; \mu$. That is, increasing temperatures favor ISB. The
parameter regions predicted by the 1LA have the same temperature
dependence although the difference in size is less significant than the
one displayed in {}Fig. 3 for the $\delta$-expansion.

{}Finally, let us investigate the possible patterns for ISB. According
to the one-loop approach (to lowest and next to leading order) as well
as to the RGA, there are only two possible phases at high $T$: either
the theory is completely symmetric or one of the two symmetries is
broken. However, including two-loop contributions and going
beyond the simple perturbative expansion could alter this picture
introducing a third possibility where the two symmetries are broken at high
$T$. To analyze this possibility one must increase the temperature
beyond the critical values shown in Tables I and II which display only
the $T_c$ connected with the phase transition from the symmetric phase
to one of the nonsymmetric phases. Surprisingly we find, that for some 
parameter
values, the mass associated with the symmetry which survives the first 
transition has a tendency to decrease as the
temperature increases beyond the first critical value. As an
example of this, we show in {}Fig. 5 the results for the case analyzed
in the first row of Table II for both $M_1^2$ and $M_2^2$. 
$M_2^2$ is positive right
above the critical temperature shown (now labeled $T_{c1}$) which is
associated with the breaking in the $\phi_1$ direction. We see that,
very quickly, $M_2^2$ becomes monotonically decreasing with the
temperature and becomes negative, through a second order phase transition,
at $T_{c2}/m_2 \simeq 18.9$. The same behavior is observed for the
parameters shown in the second row of Table II, where we observe symmetry 
breaking in $\phi_2$
direction at $T_{c2}/m_2 \simeq 139$. However, analyzing the other two
cases shown in the third and fourth rows of Table II, we find that the
would be mass associated with the second symmetry breaking increases 
monotonically
with the temperature, signalling that this symmetry possibly remains
unbroken at high temperatures. A similar behavior happens for the two
cases analyzed in Table I, for $N_1=N_2=1$. Although $M_2^2$ initially
shows a decrease in value as the temperature is raised, it soon becomes
monotonically increasing with $T$.  Note that just by interchanging the 
values of $\lambda_1$ and $\lambda_2$ one trivially observes yet another 
pattern this time with $T_{c2} < T_{c1}$. Which pattern will be actually 
followed depends on our initial choice for the values of the couplings. 

Therefore, our results suggest that a possible symmetry breaking along
the second field direction takes place, for large values of $N$, in a 
narrow region of parameters. It is possible that this
alternative symmetry breaking pattern will show up only in
nonperturbative calculations which consider up to two loop terms. It
would be interesting to investigate this possibility using other 
nonperturbative approaches.

\section{Conclusions}

We have used the optimized linear $\delta$ expansion to investigate
inverse symmetry breaking at high temperatures using multi-field
theories. Our order-$\delta^2$ calculations take full consideration of
two-loop contributions, including the momentum dependent ``setting sun''
type of diagrams. To our knowledge, a complete calculation associated
with the phenomenon of ISB/SNR which includes these contributions in the $O(N_1)\times
O(N_2)$ model has not been fully considered before. In order to asure
the reliability of the method we have started with the scalar $Z_2
\times Z_2$ model which has been extensively investigated in connection
with inverse symmetry breaking problem. We have shown that our optimized
results agree well with those obtained with the Renormalization Group
approach, especially as far as the critical temperatures are concerned.
This has allowed us to establish the $\delta$-expansion as a reliable
nonperturbative technique to investigate ISB. We have then investigated
the more realistic scalar $O(N_1) \times O(N_2)$ model which may be
related to the Kibble-Higgs sector of a $SU(5)$ grand unified model. All
our results strongly support the possibility of inverse symmetry
breaking (or symmetry nonrestoration) at high temperatures.
Surprisingly, we have also found evidence for a second phase transition
taking place for some values of the couplings and large values of $N$.
According to our results two other possible high temperature inverse symmetry
breaking patterns are

\[
O(N_1) \times O(N_2) \stackrel{T_{c1}}{\longrightarrow}
O(N_1-1) \times O(N_2) \stackrel{T_{c2}}{\longrightarrow}
O(N_1-1) \times O(N_2-1)\;,
\]

\noindent
where $T_{c1} < T_{c2}$, or
\[
O(N_1) \times O(N_2) \stackrel{T_{c2}}{\longrightarrow}
O(N_1) \times O(N_2-1) \stackrel{T_{c1}}{\longrightarrow}
O(N_1-1) \times O(N_2-1)\;,
\]

\noindent
where $T_{c1} > T_{c2}$.
  
\acknowledgements
R.O.R. is partially supported by CNPq.

\appendix

\section{}

Consider a general setting sun diagram given by ($d=4-2 \epsilon$):

\begin{equation}
G_{ijj} (p)  = \mu^{4 \epsilon} \int \frac{d^d k}{(2 \pi)^d} \int
\frac{d^d q}{(2 \pi)^d} \frac{1}{ \left[ k^2 - \Omega_i^2 + i\epsilon \right]
\left[ q^2 - \Omega_j^2 + i \epsilon  \right]  \left[  (p-k-q)^2 - \Omega_j^2 
+ i\epsilon  \right]} \;.
\label{A1}
\end{equation}

\noindent
At finite temperature we express the momentum integrals as in 
(\ref{prescription}). The discrete sums in the Matsubara frequency can
easily be done if we reexpress Eq. (\ref{A1}) in terms of the Fourier
transformed, in Euclidean time, expressions for the field propagators,

\begin{eqnarray}
G_{ijj} ({\bf p}, i \omega_n) = & & \mu^{4 \epsilon} \int 
\frac{d^{d-1} k_1}{(2 \pi)^{d-1}} \frac{d^{d-1} k_2}{(2 \pi)^{d-1}} 
\frac{d^{d-1} k_3}{(2 \pi)^{d-1}} 
\int_0^{\beta} d \tau e^{i \omega_n \tau}  G_i ({\bf k_1},\tau)
G_j ({\bf k_2},\tau) G_j ({\bf k_2},\tau) \nonumber \\
& &\delta^3 ({\bf p}-{\bf k_1}-
{\bf k_2}-{\bf k_3}) \;,
\label{A2}
\end{eqnarray}

\noindent
where $G_i ({\bf k},\tau)$ is the propagator, which
can be written in terms of a spectral function $\rho({\bf k},\eta)$, as

\begin{equation}
G_i ({\bf k},\tau) = \int_{-\infty}^{+\infty} \frac{d \eta}{2 \pi}
\left[ 1 + n(\eta) \right] \rho_i ({\bf k},\eta) e^{-\eta |\tau|} \;,
\label{A3}
\end{equation}

\noindent
where $n$ is the Bose distribution and 

\begin{equation}
\rho_i ({\bf k},\eta) = \frac{2 \pi}{2 E_i ({\bf k})} \left[
\delta(\eta - E_i ({\bf k})) - \delta(\eta + E_i ({\bf k})) \right] \;.
\label{A4}
\end{equation}
Using (\ref{A3}) in (\ref{A2}), the identities 
$e^{i \beta \omega_n}=1$,
$n(\eta) = e^{-\beta \eta} [1+ n(\eta)]$, $n(\eta) = -[1+n(-\eta)]$ and
performing the $\tau$
integration, we get

\begin{eqnarray}
\lefteqn{G_{ijj} ({\bf p}, i \omega_n)\!  =\! \mu^{4 \epsilon} \! \! \! \int 
\frac{d^{d-1} k_1}{(2 \pi)^{d-1}} \frac{d^{d-1} k_2}{(2 \pi)^{d-1}} 
\frac{d^{d-1} k_3}{(2 \pi)^{d-1}}\! 
\int_{-\infty}^{+\infty} \! \! \frac{d \eta_1}{2 \pi}
\frac{d \eta_2}{2 \pi}  \frac{d \eta_3}{2 \pi}
\rho_i ({\bf k_1}, \eta_1) \rho_j ({\bf k_2}, \eta_2) 
\rho_j ({\bf k_3}, \eta_3)} \nonumber \\
& & \times \left[1+n(\eta_1)\right] \left[1+n(\eta_2)\right] 
\left[1+n(\eta_3)\right]
\left[ \frac{1}{i\nu_n +\eta_1 + \eta_2 + \eta_3} - 
\frac{1}{i\nu_n -\eta_1 - \eta_2 - \eta_3} \right] \nonumber \\
& & \times \delta^3 ({\bf p}-{\bf k_1}-
{\bf k_2}-{\bf k_3}) \;.
\label{A5}
\end{eqnarray}

\noindent
Performing the analytic continuation in the above expression,
$i \omega_n \to p_0 + i \epsilon$, and using 

\begin{equation}
\lim_{\epsilon \to 0} \frac{\epsilon}{x^2 + \epsilon^2} = \pi \delta(x) \;,
\end{equation}

\noindent
we can separate $G_{ijj}$ in real and imaginary contributions,
$G_{ijj} = {\rm Re} G_{ijj} + i {\rm Im} G_{ijj}$. The imaginary
contribution gives the field decay rate and it is not important in the present work. The real part contributes to the thermal
mass and it is given by (after performing the $k_3$ momentum integral
with the help of the Dirac delta-function in (\ref{A5}))

\begin{eqnarray}
\lefteqn{{\rm Re} G_{ijj} ({\bf p}, p_0)  =  \int 
\frac{d^{d-1} k_1}{(2 \pi)^{d-1}} \frac{d^{d-1} k_2}{(2 \pi)^{d-1}} 
\frac{\mu^{4 \epsilon}}{8 E_i({\bf k_1})  E_j({\bf k_2})  E_j({\bf p}-{\bf k_1}-{\bf k_2}) }
\int_{-\infty}^{+\infty} d \eta_1 d \eta_3 d \eta_3 } \nonumber \\
& & \times
\left[ \delta(\eta_1 - E_i({\bf k_1})) - \delta(\eta_1 + E_i({\bf k_1}))
\right] 
\; \left[ \delta(\eta_2 - E_j({\bf k_2})) - \delta(\eta_2 + E_j({\bf k_2}))
\right] \nonumber \\
& & \times
\left[ \delta(\eta_3 - E_j({\bf p}-{\bf k_1}-{\bf k_2})) - \delta(\eta_3 
+ E_i({\bf p}-{\bf k_1}-{\bf k_2}))
\right]  \nonumber \\
& & \times 
\left[ \frac{1}{p_0 +\eta_1 + \eta_2 + \eta_3} - 
\frac{1}{p_0 -\eta_1 - \eta_2 - \eta_3} \right] 
\delta^3 ({\bf p}-{\bf k_1}-
{\bf k_2}-{\bf k_3}) \;.
\label{A6}
\end{eqnarray}

\noindent
After performing the $\eta$ integrals with the help of the Dirac 
delta-functions and after some algebra, Eq. (\ref{A6}) can be written
in terms of three terms: a temperature independent one, which is
just the zero temperature contribution $G_{ijj,0}$, 
and two other terms with one
and two Bose factors, which gives the $G_{ijj,1}$ and $G_{ijj,2}$ terms
appearing in Eq. (\ref{S13a}).

The zero temperature contributions in both cases have already been 
evaluated in the literature. The contribution $ G_{iii,0}(p)$ 
has been evaluated in details in \cite{mendels,russos}
where  the quoted result for the on mass shell ($p^2=\Omega_i^2$) case is
\begin{equation}
G_{iii,0}(p) = \frac{\Omega_i^2}{(4 \pi^2)^4}
\frac{\Gamma^2(1+\epsilon)}{(1-\epsilon)(1-2\epsilon)} 
\left ( \frac{4 \pi \mu^2}{\Omega_i^2} \right )^{2 \epsilon}
\left [ - \frac{3}{2 \epsilon^2} + \frac{1}{4\epsilon} +\frac{19}{8} \right ]
\;,
\end{equation}

\noindent
which gives

\begin{eqnarray}
G_{iii,0}(p) &=&-\frac{ 3\Omega_i^2}{2 (4 \pi)^4} 
\left[  \frac{1}{\epsilon^2} +
\frac{3-2 \gamma_E}{\epsilon} - \frac{2}{\epsilon} 
\ln\left( \frac{
\Omega_i^2}{4 \pi \mu^2} \right) \right] +
\frac{p^2}{4 (4 \pi)^4 \epsilon} \nonumber \\
&-& \frac{3 \Omega_i^2}{ (4 \pi)^4} 
\left[ \ln^2 \left( \frac{
\Omega_i^2}{4 \pi \mu^2} \right) + \left(2 \gamma_E -\frac{17}{6}\right) 
\ln\left(
\frac{
\Omega_i^2}{4 \pi \mu^2} \right) +  1.9785 \right]\;,
\label{div7}
\end{eqnarray}
where we purposefully left the momentum dependence in the relevant
divergent term
to make explicit the need for a wave-function renormalization counterterm.

The zero temperature contribution to the mixed setting sun diagram is also 
given in Ref. \cite {russos}, for the on shell case ($p^2=\Omega_i^2$) can be written as
\begin{eqnarray}
\lefteqn{G_{ijj,0}(p) = \frac{\Omega_j^2}{(4 \pi)^4}
\frac{\Gamma^2(1+\epsilon)}{(1-\epsilon)(1-2\epsilon)} 
\left ( \frac{4 \pi \mu^2}{\Omega_j^2} \right )^{2 \epsilon}
\left \{ - \frac{2+n^2}{2 \epsilon^2} + \frac{n^2}{\epsilon}
\left [ \frac{1}{4} + \ln (n^2) \right ] \right .} \nonumber \\
& & + \left . \left [ 1 + \frac{11}{8}n^2 - \left (1 +\frac{n^2}{2} 
\right )\ln(n^2) - \frac {1}{2} n^2 \ln^2 (n^2) + 
\frac {(1-n^2)^2}{n^2}\left ( {\rm Li}_2(1-n^2) - \frac{\pi^2}{6}\right ) 
\right ]\right \} \;,
\end{eqnarray}
where $n=\Omega_i/\Omega_j$ and 
${\rm Li}_2 (z) = \sum_{l=1}^{\infty} z^l/l^2$. 
We then obtain the result 
\begin{eqnarray}
G_{ijj,0}(p) &=& - \frac{\Omega_j^2}{(4\pi)^4} 
\left \{ \frac{1}{\epsilon^2}
\left (1 +\frac {n^2}{2} \right ) 
+ \frac {1}{\epsilon} 
\left ( 1+\frac {n^2}{2} \right ) \left [ 3 - 2 \gamma_E - 2 \ln 
\left ( \frac {\Omega_j^2}{4\pi \mu^2} \right ) \right ] \right . \nonumber\\
&-& \left .\frac{p^2}{4 \Omega_j^2 \epsilon} - \frac{n^2}{\epsilon} \ln (n^2)  
\right \} \nonumber \\
&-&\frac{\Omega_j^2}{(4\pi)^4}  \left( 1 +\frac {n^2}{2} \right ) 
\left [ 7 + \frac{\pi^2}{6} - 6 \gamma_E + 2 \gamma^2_E - 2(3 - 2 \gamma_E)
\ln \left ( \frac {\Omega_j^2}{4\pi \mu^2} \right ) + 2 \ln^2 \left ( 
\frac {\Omega_j^2}{4\pi \mu^2} \right ) \right ] \nonumber \\
&+&\frac{\Omega_j^2}{(4\pi)^4}\left [ 1 + \frac {11}{8}n^2 - \left ( 1 + 
\frac{n^2}{2} \right ) \ln(n^2) - \frac{1}{2} n^2 \ln^2(n^2)\right . \nonumber \\
 &+& \left . 
\frac {(1-n^2)^2}{n^2} \left ( {\rm Li}_2(1-n^2) - \frac{\pi^2}{6} 
\right ) \right ]
\nonumber \\
&+&\frac{\Omega_i^2}{(4\pi)^4}\left \{ \left [ 3 - 2 \gamma_E - 
2 \ln\left ( \frac {\Omega_j^2}{4\pi \mu^2} \right ) \right ] 
\left ( \frac {1}{4} + \ln (n^2) \right ) \right \} \;.
\end{eqnarray}

The finite temperature terms $G_{ijj,1}$ and $G_{ijj,2}$ can be
worked out as follows.
Taking ${\bf p} =0$ in these terms, allows us to reexpress them by
(using ${\bf k}_1 . {\bf k}_2 = k_1 k_2 \cos \theta$)

\begin{eqnarray}
\lefteqn{{\rm Re} G_{ijj,1} ({\bf 0},p_0) =
\mu^{4 \epsilon} \int 
\frac{d^{d-1} k_1}{(2 \pi)^{d-1}} \frac{d^{d-1} k_2}{(2 \pi)^{d-1}} 
\frac{1}{8 E_i({\bf k_1})  E_j({\bf k_2})} \frac{1}{k_1 k_2}    }
\nonumber \\
& & \times \left[ 
n (E_i({\bf k}_1))
\frac{\partial}{\partial \cos \theta} \ln \left\{
\left[p_0^2 - \left( E_i({\bf k_1}) +  E_j({\bf k_2}) +  
E_j({\bf k_1}+{\bf k_2}) \right)^2 \right] \right. \right. \nonumber \\
& & \left. \left. \hspace{4cm}\times 
\left[p_0^2 - \left(- E_i({\bf k_1}) +  E_j({\bf k_2}) +  
E_j({\bf k_1}+{\bf k_2}) \right)^2 \right] \right\} \right. \nonumber \\
& & \left. + 
2 n (E_j({\bf k}_2))
\frac{\partial}{\partial \cos \theta} \ln \left\{
\left[p_0^2 - \left( E_i({\bf k_1}) +  E_j({\bf k_2}) +  
E_j({\bf k_1}+{\bf k_2}) \right)^2 \right] \right. \right. \nonumber \\
& & \left. \left. \hspace{4cm} \times
\left[p_0^2 - \left(E_i({\bf k_1}) -  E_j({\bf k_2}) +  
E_j({\bf k_1}+{\bf k_2}) \right)^2 \right] \right\} \right]
\label{A7}
\end{eqnarray}

\noindent
and

\begin{eqnarray}
{\rm Re} G_{ijj,2} ({\bf 0},p_0) =& &
\mu^{4 \epsilon} \int 
\frac{d^{d-1} k_1}{(2 \pi)^{d-1}} \frac{d^{d-1} k_2}{(2 \pi)^{d-1}} 
\frac{1}{8 E_i({\bf k_1})  E_j({\bf k_2})} \frac{1}{k_1 k_2} 
\nonumber \\
& & \times \left[ 
2 n (E_i({\bf k}_1)) n (E_j({\bf k}_2)) 
+ n (E_j({\bf k}_1)) n (E_j({\bf k}_1+{\bf k}_2)) \right] \nonumber \\
& & \times \frac{\partial}{\partial \cos \theta} \ln \left\{
\left[p_0^2 - \left( E_i({\bf k_1}) +  E_j({\bf k_2}) +  
E_j({\bf k_1}+{\bf k_2}) \right)^2 \right] \right. \nonumber \\
& & \left. \hspace{2cm}\times 
\left[p_0^2 - \left(- E_i({\bf k_1}) +  E_j({\bf k_2}) +  
E_j({\bf k_1}+{\bf k_2}) \right)^2 \right] 
\right. \nonumber \\
& & \left.  \hspace{2cm}\times 
\left[p_0^2 - \left(E_i({\bf k_1}) - E_j({\bf k_2}) +  
E_j({\bf k_1}+{\bf k_2}) \right)^2 \right]
\right. \nonumber \\
& & \left.  \hspace{2cm}\times 
\left[p_0^2 - \left( E_i({\bf k_1}) +  E_j({\bf k_2}) - 
E_j({\bf k_1}+{\bf k_2}) \right)^2 \right]
\right\} \;.
\label{A8}
\end{eqnarray}

\noindent
{}For $i=j$ (equal mass propagators) Eqs. (\ref{A7}) and (\ref{A8})
give the same expressions obtained by Parwani in Ref. \cite{Parwani}.
In special, we note that, in Eq. (\ref{A7}), the terms given by
$\partial/\partial \cos \theta {\rm ln} \{ \ldots \}
\to 2 k_1/k_2$ as $k_2 \to \infty$. 
We can subtract and add this term in the appropriate places in (\ref{A7}),
obtaining the analogous expressions 
given by Parwani,

\begin{equation}
-\frac{\lambda^2 \delta^2 N_j}{2} 
{\rm Re} G_{ijj,1} ({\bf 0},p_0) = F_{ijj,0} + F_{ijj,1} + F_{ijj,2} (p_0)
\;,
\label{FFF}
\end{equation}

\noindent
where

\begin{equation}
{}F_{ijj,0} = - \delta^2 N_j \frac{\lambda^2 T^2}{ (4\pi)^2} 
\frac{1}{\epsilon}\left[ h\left(\frac{\Omega_i}{T}\right) +
2 h\left(\frac{\Omega_j}{T}\right) \right]
\label{Fijj0}
\end{equation}
and
\begin{eqnarray}
{}F_{ijj,1} &=& - \delta^2 N_j \frac{\lambda^2 T^2}{ (4 \pi)^2} 
\left\{
h\left(  \frac{\Omega_i}{T}\right)
\left [ - \ln\left(
\frac{
\Omega_j^2}{4 \pi \mu^2} \right) +2 - \gamma_E \right ] \right.
\nonumber \\
&+& \left. 2 h\left(  \frac{\Omega_j}{T}\right)
\left [ - \ln\left(
\frac{
\Omega_i^2}{4 \pi \mu^2} \right) +2 - \gamma_E \right ] \right\} \;,
\label{Fijj1}
\end{eqnarray}

\noindent
where in the above equations, $h(y_i)$ is given by (\ref{hy}).
The remaining terms, $F_{ijj,2}$ and $G_{ijj,2}$ can be evaluated
on-shell and a similar contribution have already been computed earlier 
in the literature, see Ref. \cite{arnold}, from where we obtain 
\begin{equation}
{}F_{ijj,2} (\Omega_i) - 
\frac{\lambda^2 \delta^2 N_j}{2} 
{\rm Re} G_{ijj,1} ({\bf 0}, \Omega_i) \sim
- \delta^2 N_j^2 \frac{\lambda^2 T^2}{8 (4 \pi)^2}
\left[ 2 \ln \left( \frac{\Omega_i + 2 \Omega_j}{3 T}\right) + 5.0669 \right]
\;.
\label{F2G2}
\end{equation}

The finite temperature contributions $G_{iii,1}$ and $G_{iii,2}$ are given, as
in Ref.~ \cite {Parwani} and, from the previous equations, 
Eqs. (\ref{FFF}) - (\ref{F2G2}), they can be written as

\begin{equation}
- \delta^2 \frac{\lambda_i^2 (N_i+2)}{18} 
{\rm Re} [G_{iii,1} ({\bf 0}, \Omega_i)]  = F_{iii,0} + 
F_{iii,1} + F_{iii,2} \;,
\end{equation}

\noindent
where\footnote{Note that there is a misprint
in Eq. (2.32) of Ref. \cite{MR}, where there is an extra $1/2$ multiplying that equation.}

\begin{equation}
F_{iii,0} = - \delta^2 \frac{\lambda_i^2 (N_i+2) T^2}{3 (4\pi)^2} 
\frac{1}{\epsilon} h\left(\frac{\Omega_i}{T} \right) 
\;,
\label{div8}
\end{equation}

\begin{equation}
F_{iii,1} = - \delta^2\frac{\lambda_i^2 (N_i+2) T^2}{3 (4 \pi)^2} 
h\left(\frac{\Omega_i}{T}\right) \left [ - \ln\left(
\frac{
\Omega_i^2}{4 \pi \mu^2} \right) +2 - \gamma_E \right ]
\label{F_1}
\end{equation}

\noindent
and

\begin{equation}
F_{iii,2}(\Omega_i) -\frac{\delta^2\lambda_i^2 (N_i+2)}{18} 
{\rm Re}[G_2 ({\bf 0},
\Omega_i)] \sim 
\delta^2  \frac{\lambda_i^2 (N_i+2) T^2}{72 (4 \pi)^2} 
\left[
\ln \left(\frac{\Omega_i^2}{T^2}\right) + 5.0669 \right]\;.
\end{equation}

Putting all these contributions together, we obtain the results
given in Eqs. (\ref{S12}) and (\ref{S13}).

\section{}

To obtain the total finite order $\delta^2$ contribution one can add all
divergences appearing in Eqs.
(\ref{S2})-(\ref{S9}), (\ref{S10}), (\ref{S11}),
(\ref{S12}) and (\ref{S13}).
As it can be easily seem all the temperature dependent divergences cancel 
exactly and one is left with temperature independent poles.

By looking at all terms which contribute to this order one can 
identify two classes. The first is composed by diagrams such as the 
ones described by Eqs. (\ref{S4}),(\ref{S5}),(\ref{S6}),(\ref{S7}),
(\ref{S8}),(\ref{S9}),(\ref{S10}),(\ref{S11}),(\ref{S12}) and 
(\ref{S13}). All of them are analogous to the diagrams which 
appear at second order in the couplings in the original theory and can be 
rendered 
finite by similar mass and wave-function counterterms. This procedure has 
already been illustrated at order-$\delta$. One can generalize this procedure 
by stating that  diagrams belonging to a general order-$\delta^n$, and containing any 
combination $\lambda^m \lambda_1^k \lambda_2^l$ such that $m+k+l=n$ will 
be renormalized exactly as when ordinary perturbation theory is applied 
to the original theory.   Then, one just has to replace the original masses 
with the relevant interpolating masses $\Omega_{1,2}$.  It is easy to check 
that for those diagrams
the most divergent terms will display $\epsilon^{-n}$ poles.

The second kind of diagram is exclusive of the interpolated theory and 
carries at least one $\delta \eta_i^2$ (or $\delta \eta_j^2$) vertex. At ${\cal O}(\delta^2)$ these
diagrams are described by Eqs. (\ref{S2}) and (\ref{S3}),
which display the divergent $\eta^2_i$ and $\eta^2_j$ pieces. 
 Looking at 
${\cal L}^{\delta}_{\rm ct}$ one identifies a $\eta_i^2$,$\eta^2_j$-dependent 
coefficient whose Feynman rule is $i \delta B_i^{\delta} (\eta_1,\eta_2)$. 
Since the actual pole is of order-$\delta^2$ one identifies 
this coefficient as having the same structure as the mass counterterm $B^{\delta^1}(\Omega_1,\Omega_2)$, displayed in Eq. (\ref{S1}), except that we now have $\eta^2_{i,j}$ instead of $\Omega^2_{i,j}$.
Therefore, ${\cal O}(\delta^n)$ diagrams belonging to the second class will
make use of the counterterm $\delta B^{\delta^n} (\eta_1,\eta_2)$. This coefficient is similar  to 
$B^{\delta^{1-n}}(\Omega_1,\Omega_2)$ which has been evaluated in a previous order. One can also
easily check that for these diagrams the most divergent terms will have
$\epsilon^{n-1}$ poles. Moreover, power counting reveals that those
$\delta \eta_i^2$ (or $\delta \eta_j$) insertions make the loops more convergent. {}For example,
{\it all} one loop diagrams of order ${\cal O}(\delta^n)$, 
with $n \ge 3$ are
finite. 

{}The renormalization prescription adopted
here is analogous to the one shown in \cite{MR} for
the one-field case,
where we have shown that the order by order renormalization holds at 
any higher orders in $\delta$. In Ref. \cite{MR} the renormalization procedure is treated in more detail for the simple $\lambda \phi^4$ case.


\begin{table}
\caption{Results for the critical temperature as obtained in the
1-loop approximation, the renormalization group approach in 
Ref. [14] and in the $\delta$-expansion, respectively.}
\begin{tabular}{c|c|c|c|c|c}
$\lambda_1$ & $\lambda_2$ &  $\lambda$ & $T_c^{\rm 1-loop}/m_1$ & 
$T_c^{\rm RGA}/m_1$& $T_c^{\delta^2}/m_1$ \\ 
\hline
0.06   &  1.8 & - 0.1   & 24.5  & 33.6  & 33.3 \\
0.0167 &  0.6 & - 0.025 & 53.7  & 67.9  & 66.9 \\       
\end{tabular}
\end{table}

\vspace{2.1cm}

\begin{table}
\caption{Results for the critical temperature in the $O(90) \times O(24)$
scalar model.}
\begin{tabular}{c|c|c|c|c}
$\lambda_1$ & $\lambda_2$ &  $\lambda$ &  $T_c^{\rm 1-loop}/m_1$ & 
$T_c^{\delta^2}/m_1$ \\ 
\hline 
0.06   &  1.8 & - 0.1   &  6.5  & 15.3  \\
0.0167 &  0.6 & - 0.025 & 16.5  & 65.4  \\ 
0.9    &  1.0 & - 0.141 &  2.2  & 5.0  \\
0.8    &  0.7 & - 0.091 &  3.4  & 10.0  \\
\end{tabular}
\end{table}

\newpage

\begin{center}
{\large \bf Figure Captions}
\end{center}

{\bf Figure 1:} The critical temperature $T_c/m_1$ as a function of 
$\lambda$ ($<0$) for $N_1=1$ and $N_2=1$ with the following parameter
values: $m_{1,2}^2/\mu^2 =1.0$, $\lambda_1=0.018$ and $\lambda_2=0.6$. The
dashed line is the one-loop prediction, the continuous line is our
result and the dots represent the values obtained with the RGA.
\\

{\bf Figure 2:} Region of ISB at fixed $\lambda_1=0.8$ for $N_1=90$ and
$N_2=24$. The thick parabola limits the region for which the potential
is bounded. The continuous and dashed lines are the zeroth order 
(at $T/\mu=5.0$) and
first order (at an arbitrarily large $T$) results given in Ref. \cite{gap}. 
The dot-dashed line is our
result at $T/\mu = 5.0$. The region of ISB is the one in between the
boundness curve and the other curves for each case.
\\

{\bf Figure 3:} Same as in {}Fig. 2, but now at fixed $\lambda_2=1.0$.
\\

{\bf Figure 4:} The region of ISB for two different temperatures 
($\lambda_1=0.8$):
$T/\mu=5.0$ (dashed line) and $T/\mu=10.0$ (thin full line). 
\\

{\bf Figure 5:} The behavior of $M_1^2$ (dashed line) and $M_2^2$ (upper
full line) as a function of the temperature for $\lambda_1=0.06$
$\lambda_2 =1.8$ and $\lambda = - 0.1$ ($N_1 = 90$ and $N_2 = 24$).

\begin{figure}[b]
\epsfysize=15cm 
{\centerline{\epsfbox{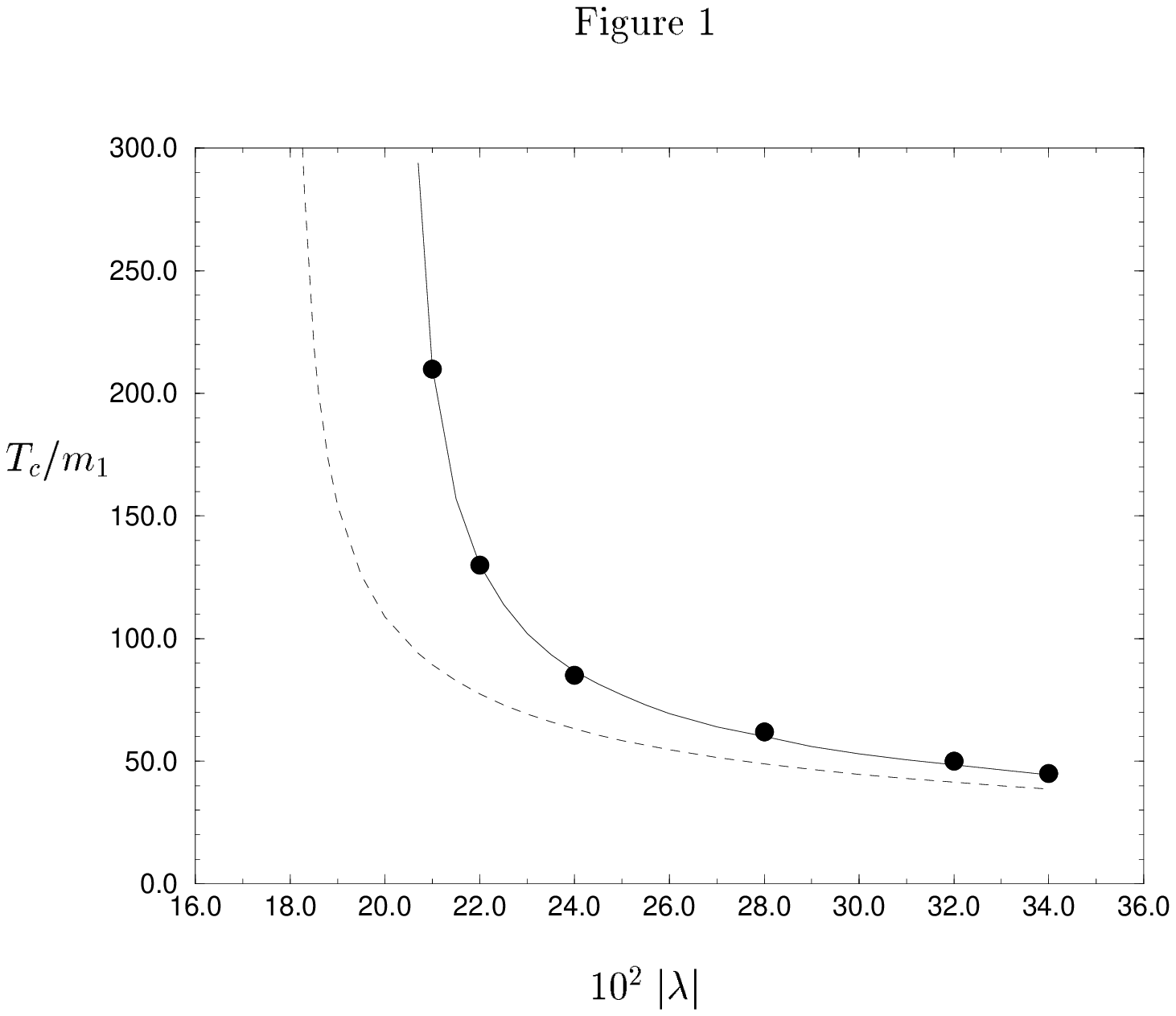}}}

\vspace{1cm}

\end{figure}

\newpage

\begin{figure}[b]
\epsfysize=15cm 
{\centerline{\epsfbox{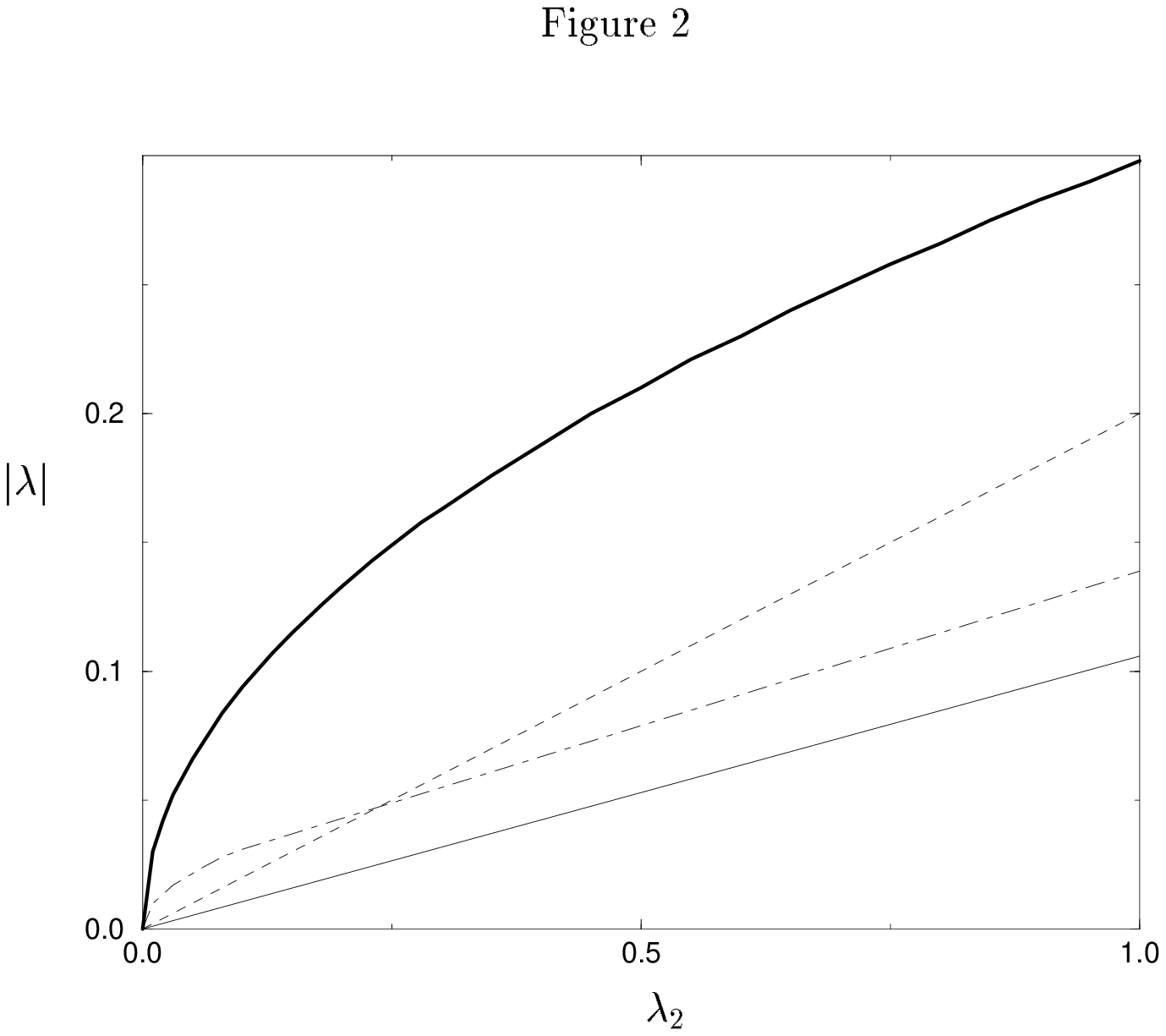}}}

\end{figure}

\newpage

\begin{figure}[b]
\epsfysize=15cm 
{\centerline{\epsfbox{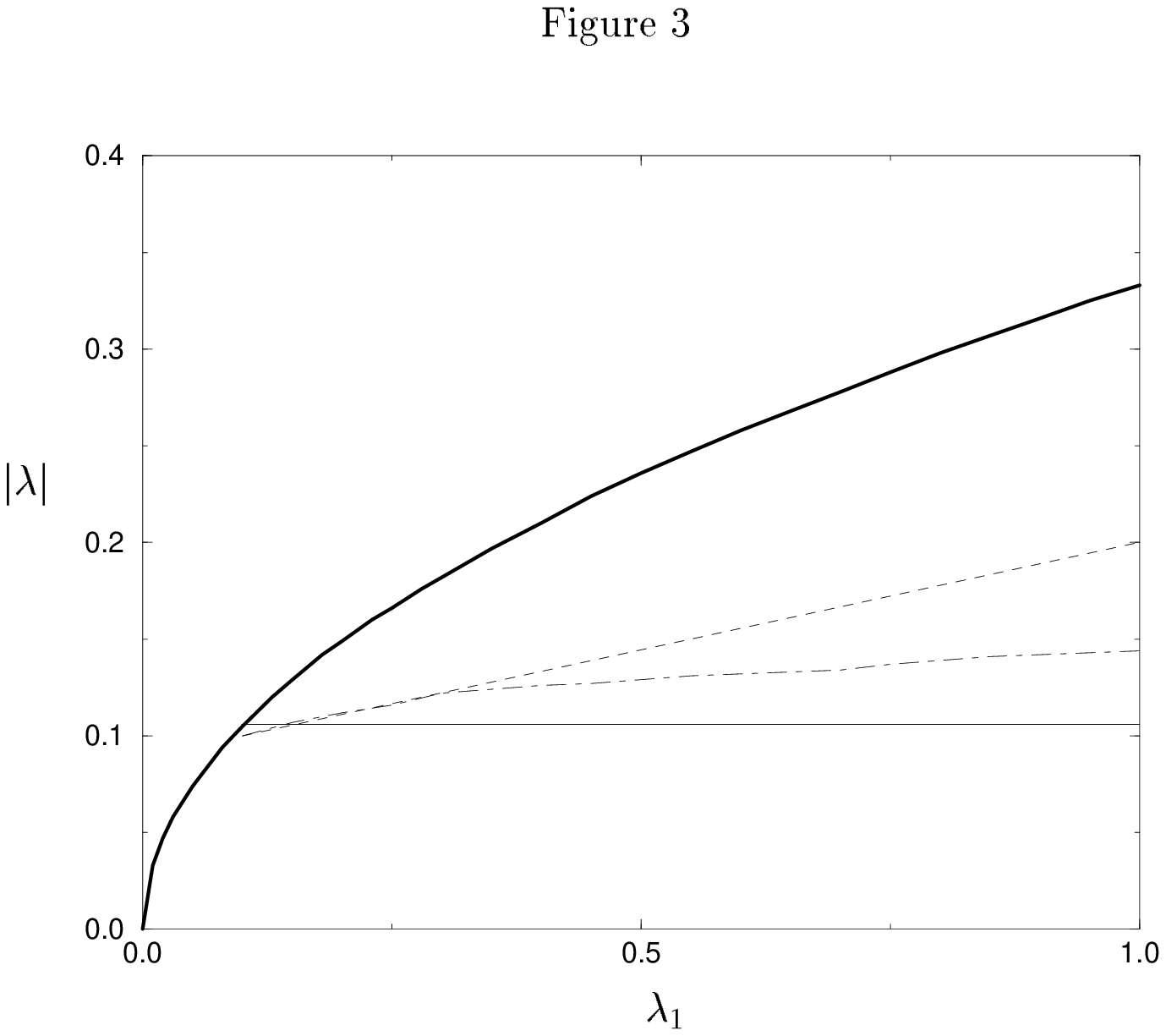}}}
\end{figure}

\newpage

\begin{figure}[b]
\epsfysize=15cm 
{\centerline{\epsfbox{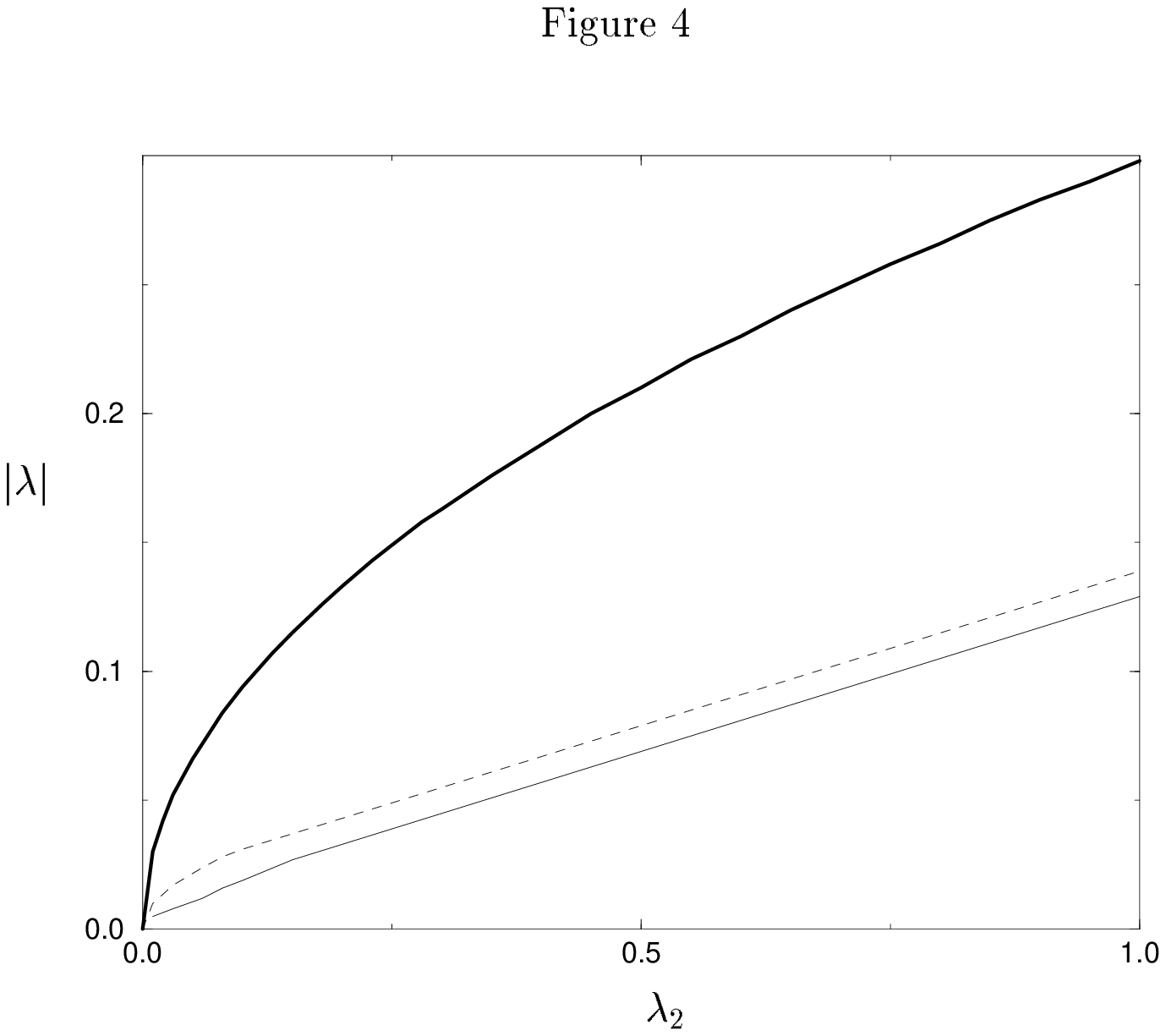}}}
\end{figure}

\newpage

\begin{figure}[b]
\epsfysize=14cm 
{\centerline{\epsfbox{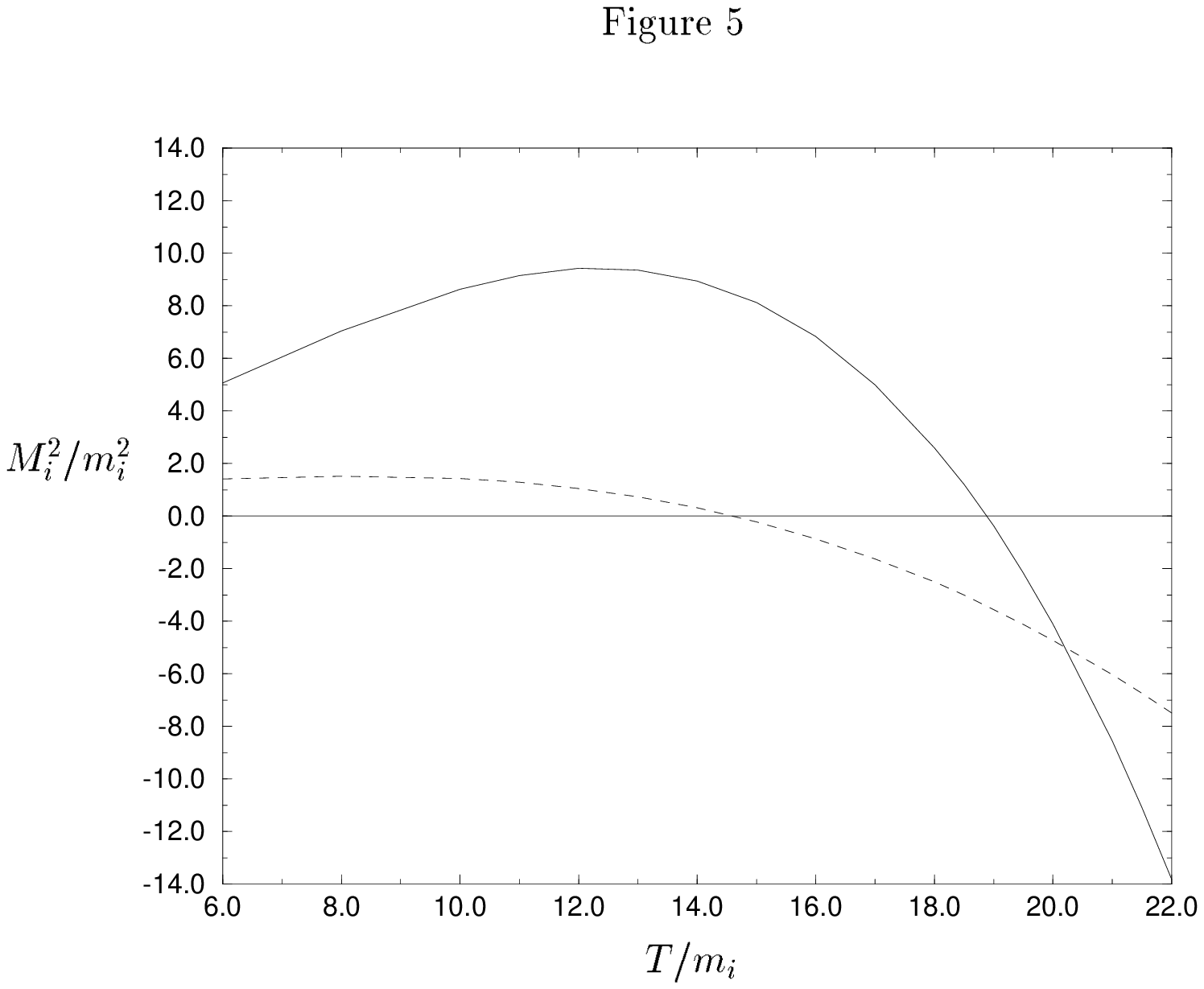}}}
\end{figure}

\end{document}